\documentclass{article}

\usepackage{psfrag,latexsym,natbib}
\usepackage{graphicx}
\renewcommand{\vec}[1]{\mbox{\boldmath $#1$}} 

\def\bra{\langle}
\def\ket{\rangle}
\def\p{\partial}

\def\beq{\begin{equation}}
\def\eeq{\end{equation}}
\def\la{\label}

\def\r#1{(\ref{#1})}

\def\diss{\varepsilon}
\def\tg{\widetilde{g}}
\def\spacce#1{\hskip #1pt}
\def\drawline#1#2{\raise 2.5pt\vbox{\hrule width #1pt height #2pt}}
\def\solid{\drawline{24}{.5}\nobreak}

\def\bdash{\hbox{\drawline{5.8}{.5}\spacce{2}}}

\def\dashed{\bdash\bdash\bdash\nobreak}

\def\chndot{\hbox%
{\drawline{4.6}{.5}\spacce{2}\drawline{1}{.5}\spacce{2}\drawline{4.6}{.5}\spacce{2}\drawline{1}{.5}\spacce{2}\drawline{4.6}{.5}}\nobreak }
\def\circle{$\circ$\nobreak }

\def\trian{\raise 1.25pt\hbox{$\scriptstyle\triangle$}\nobreak}

\def\dtrian{\raise 1.25pt\hbox%
{$\scriptscriptstyle\bigtriangledown$}\nobreak}

\def\squar{\raise 1.25pt\hbox{$\scriptstyle\Box$}\nobreak}

\def\diamon{\raise 1.25pt\hbox{$\scriptstyle\diamond$}\nobreak}

\newcommand{\soliddtrian}{$\blacktriangledown$\nobreak}

\def\linedtri1{\hbox{\bdash\hspace{-1.6mm}$\bigtriangleup$\hspace{-0.8mm}\bdash}\nobreak}
\def\soliddtrian1{$\blacktriangledown$\nobreak}
\def\solidrtrian2{$\blacktriangleright$\nobreak}
\def\solidltrian3{$\blacktriangleleft$\nobreak}

\title{\huge\bf Machine-aided turbulence theory}

\author{{\bf Javier Jim\'enez}\\
{\rm School of Aeronautics, U. Polit\'ecnica Madrid, 28040 Madrid Spain}}
\date{\today}

\begin{document}
\maketitle
\begin{abstract}
The question of whether significant sub-volumes of a turbulent flow can be identified by
automatic means, independently of a-priori assumptions, is addressed using the example of
two-dimensional decaying turbulence. Significance is defined as influence on the future
evolution of the flow, and the problem is cast as an unsupervised machine `game' in which
the rules are the Navier--Stokes equations. It is shown that significance is an intermittent
quantity in this particular flow, and that, in accordance with previous intuition,
its most significant features are vortices, while the least significant ones are
dominated by strain. Subject to cost considerations, the method should be applicable to more
general turbulent flows.
\end{abstract}


\section{Introduction}

It may appear that the continuous vector fields of fluid flow are not good candidates for
the identification of individual structures, but the search for features that explain a
substantial fraction of the flow dynamics while covering a relatively small fraction of the
volume has been a recurrent theme in fluid mechanics. We intuitively recognise waves in the
sea, eddies in rivers, and clouds in the sky. The identification is harder in the chaotic
velocity distributions of turbulent flows, but even there the description of the dynamics in
terms of `eddies' began early \citep{rich20}, and accelerated with the advent of
numerical simulations. The structural view of turbulent shear flows has recently been
reviewed by \cite{brownr:12} away from walls, and by \cite{jim18} near them.
Curiously, the structures of the nominally simpler case of homogeneous turbulence have been
investigated less, and, while vortices have been considered often \citep{vinc91,jwsr}, there
is clear evidence of larger kinetic-energy structures \citep{cardesa17} which are not
necessarily related to the vorticity.

However, the way in which these structures were initially identified deserves some
discussion. Consider the case of wall turbulence. Its best-known structures are streaks and
quadrants. Both are known to be dynamically significant \citep{jim18}, but neither was
originally identified from dynamical considerations. Streaks were educed from the behaviour
of scalar tracers \citep{kline67}, and quadrants were proposed because they are intense
concentrations of a quantity, the tangential Reynolds stress, whose mean value is
dynamically significant \citep{lu:wil:73}. Both approaches are valid, but they raise the
question of whether these really are the most significant structures in the flow, and even
whether we may be missing other dynamically relevant features. 

The purpose of this paper is to discuss whether it is possible to decide, at least in principle, if some
parts of the flow are more significant than others, and to explore means of identifying
them. We should first distinguish between representativeness and significance. Most of the
examples mentioned above are structures chosen as representative of some property, in the
sense, for example, that flow regions with a particularly high kinetic energy may be
considered representative of the kinetic energy of the flow. This is connected with the
concept of intermittency, in which a substantial fraction of some intensive quantity is
concentrated in a small fraction of the volume. Intermittency is probably related to the
relevance of those regions to the dynamics, but the two concepts are different. For example,
it was noted by \cite{jim18} that the intermittency of a variable depends on how it is
processed. The modulus of the velocity is much less intermittent than its square (i.e., the
kinetic energy), even if both represent the same physical quantity. We define
significance as the influence of a feature on the future evolution of the flow.
Specifically, after preparing an initial condition, we define significant subsets with
a given volume as those whose modification produces a large divergence in the flow behaviour
after some time.

Inspired by the recent proliferation of large-scale data analyses, we discuss a strategy
of `blindly' choosing subsets of the flow, determining their significance, and examining the
properties of those found to be most significant.

This approach is not entirely novel. Modifying parts of a system and observing the
consequences is a classical approach in the natural sciences. An early example in free-shear
flows is \cite{cimbala:88}, who artificially damped the coherent structures of a
turbulent wake to test whether they were due to a local instability
or leftovers from transition. In wall turbulence, \cite{jpin} modified different terms of
the evolution equations in simulations of turbulent channels to observe their effect.
Both cases can be considered examples of `a posteriori' testing, in which structures 
believed to be significant are tested to be so. In contrast, the procedure in
this paper could be considered `a priori', in the sense that parts of the flow are tested
blindly and, if especially significant ones are identified, their properties are examined.

Our approach is a variant of unsupervised reinforcement machine learning, although it will
become clear that part of the learning loop is still reserved to the human researcher. In
the more common supervised approach, an algorithm is trained with samples of a known
pattern, e.g. pictures, and asked to recognise that pattern in subsequent cases. In
unsupervised learning there is no training set, and the algorithm is expected to identify
patterns `without guidance'. Even in this case, a goal has to be specified. A common
approach is to define a `correct' answer in the form of a reward, and instruct the algorithm
to maximise it. The best known examples are strategy games, such as chess or Go
\citep{AlphaGo17}, in which the winning criterion is part of the rules, and the machine is
instructed to improve its strategy by playing against itself. In our case, the correct
answer is codified in the similarity of two flow fields, one of which is defined to be
correct, and the rules of the `game' are the equations of motion.

The main difficulty of this programme is cost, because many candidate subsets have to be
tested, and the procedure is only useful if it is applied to enough flow fields to
permit meaningful statistics. Assume for example that the flow field is divided into $N$
cells, and that subsets involve $n<N$ cells. The number of possible combinations is the
binomial coefficient ${N\choose n}$, which soon becomes impractically large. Luckily, the
tests on different cases are independent of one another, and, depending on the particular
search algorithm employed, the same is true for individual subsets. This means that they can
be trivially parallelised in multiprocessor computers.

Some simplifications are still required for the sake of economy, and we use an approximate
search related to the `random forest' schemes used in automatic classification
\citep{Breiman:01}. However, our emphasis is not on optimising the search algorithm, for
which many references exist \citep{LecunEtal2015}, but on testing whether the machine-aided
learning strategy gives reasonable answers in the context of flow evolution. To make the
problem tractable, we use the example of homogeneous two-dimensional turbulence in a
periodic box, whose simulation is reasonably cheap. Its significant structures, compact
vortices, are also believed to be well understood \citep{mcwilliams90b}, providing a
convenient check for the performance of the algorithm. Section \ref{sec:ploff} details the
identification algorithm. Its application to two-dimensional turbulence is discussed in
\S\ref{sec:tur2d}, and \S\ref{sec:conc} concludes. Algorithmic and
performance details are deferred to appendix \ref{app:search}.

\section{Proper labelling}\la{sec:ploff} 

The result of applying the above strategy to a given flow field is its segmentation into
subsets labelled as more or less significant for some particular application, and
we will refer to labellings that are optimal in this sense as `proper'. Consider the
following algorithm:
\vspace{1ex}%
\noindent{}%
\newcounter{plc}
\begin{list}{(PL\arabic{plc}):\ }{\usecounter{plc}\setlength{\parsep}{0ex}\setlength{\topsep}{0ex}}%

\item\la{ploff:1} Construct a reference initial flow field, $g_{ref}(\vec{x},t=t_{ref})$,
and choose a target time, $t=t_{ref}+T$.
\item\la{ploff:2} Create a test field $\tg$ by substituting $g_{ref}$ by something else over
a subset $A$ covering a fixed fraction of the domain volume, e.g. $\tg(A,t=t_{ref})=0$.
\item\la{ploff:3} Evolve both $g_{ref}$ and $\tg$ to $t=t_{ref}+T$, using the equations of motion.
\item\la{ploff:4} Iterate over $A$, and label as most significant the subset maximising (or
minimising) the magnitude of  the resulting perturbation, $\diss(T)=\|\tg(t_{ref}+T)-g_{ref}(t_{ref}+T)\|$.
\end{list}\vspace{1ex}
%
This definition is ambiguous in several respects, the most obvious being the
freedom of how to modify the test field in step PL\ref{ploff:2}. This freedom
is not absolute, because the resulting flow field has to satisfy, for example, continuity,
but the way in which such constraints are enforced only adds choices to the 
procedure. Another choice is the perturbation norm in PL\ref{ploff:4}. We use in this
paper the $L_2$ norm of the velocity field, but enstrophy or some measure of
scalar mixing could be equally valid, and would probably give different results.

Another question is whether to consider as significant the subsets resulting in the maximum
or in the minimum deviation. Both cases are interesting, because one of the questions to be
answered is whether there really exist parts of the flow that are more significant than
others. The ratio between the two extreme deviations, $R_\diss=\diss_{max}/\diss_{min}$, is
a figure of merit that characterises the existence of significant structures, and therefore
the coherence properties of the flow field as a whole. Finally, the above algorithm labels a
single reference initial condition. Most interesting questions depend on applying it to many
such cases, and abstracting from the results the statistical characteristics that define
significant subsets.

The above procedure is related to the Lyapunov analysis of dynamical systems, whose
aim is to identify the maximum growth rate of infinitesimal perturbations. However, it
differs from it in three important aspects. In the first place, our perturbations are not
infinitesimal, since they are intended to model the nonlinear evolution of turbulence. The
mathematical theory for finite-amplitude perturbations is not as developed as for
infinitesimal ones \citep{EckRue:85}, but they have been used to
characterise predictability \citep{CencVulp:13}. Secondly, Lyapunov analysis searches for
properties of the dynamical system in the long-time limit, while we are interested in
labelling significant regions of individual realisations after a finite evolution time. Any
information is thus linked to a particular flow field and to a given time horizon, and
generic properties only arise after ensemble averaging over many realisations
\citep{DingLi:07}. Lastly, and most importantly, Lyapunov analysis seeks to identify optimal
perturbations, while we restrict ourselves to perturbations within the prescribed support of
the subset $A$. This is required to represent individual structures, of which several may
exist in any given flow realisation, but the perturbation growth is necessarily suboptimal.

\begin{figure}
\vspace*{8mm}%
\centerline{\includegraphics[width=.95\textwidth,clip]{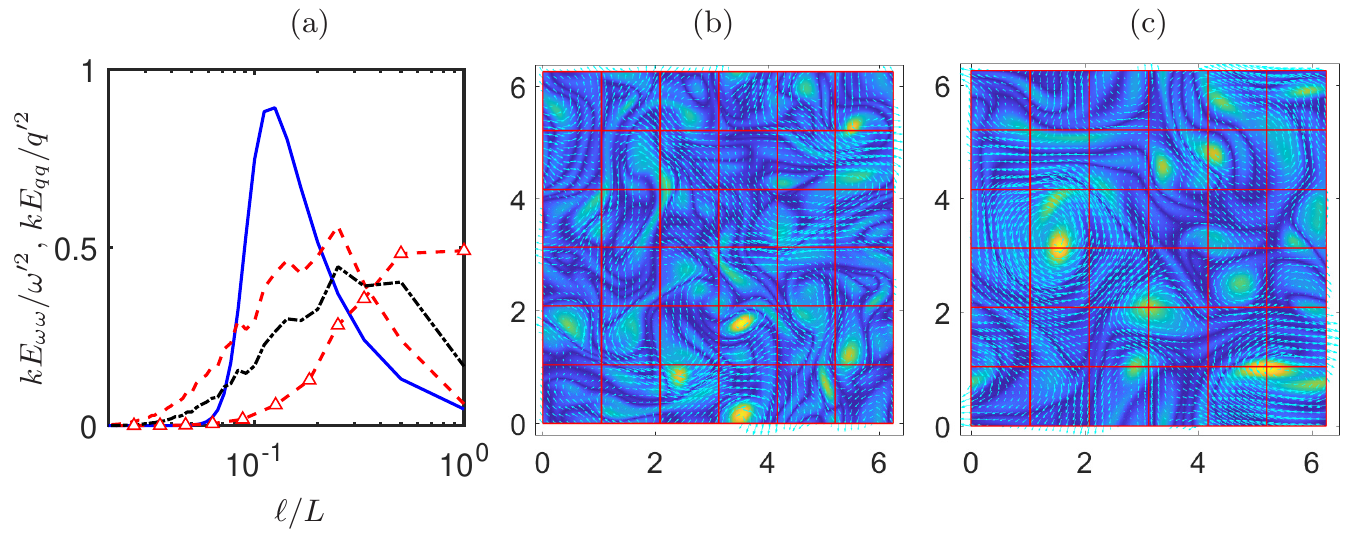}}%
%
%
\caption{%
(a) Premultiplied vorticity spectra of the simulations at the two times used as references.
The wavelength, $\ell=2\pi/k$, is defined from the wavenumber magnitude $k$. \solid, $t=0$;
\dashed, $\omega'_0 t= 11$; \chndot, $\omega'_0 t=26$. The line with symbols is the
premultiplied energy spectrum at $\omega'_0 t= 11$.
(b) Velocity and vorticity magnitude of the reference field at $\omega'_0 t=11$. The
overlaid grid is used to select the test subsets.
(c) As in (b), for $\omega'_0 t=26$. 
}
\label{fig:tur2d}
\end{figure}

\section{Application to two-dimensional turbulence}\la{sec:tur2d}

To illustrate the procedure, we apply it to the relatively simple case of identifying the
most and the least significant features of decaying  isotropic two-dimensional turbulence.
Simulations are performed in a doubly periodic square box of side $L=2\pi$, using a standard
spectral Fourier code dealiased by the 3/2 rule. Time advance is third-order Runge-Kutta.
The flow is defined by its velocity field $\vec{u}=(u,v)$ as a function of
$\vec{x}=(x,y)$. The one-component vorticity is $\omega=\p_x v - \p_y u$, and the rate
of strain tensor is $s_{ij}=(\p_iu_j+\p_ju_i)/2$, where the subindices of the
partial derivatives range over $(x,y)$, and those of the velocity components over $(u,v)$.
Its magnitude is defined by $S^2=s_{ij}s_{ij}$, where repeated indices imply summation. Natural
time and velocity scales are, respectively, the root-mean-square vorticity magnitude
$\omega' =\bra \omega^2\ket ^{1/2}$, where the time-dependent average $\bra\cdot\ket$ is
taken over the full computational box, and $q' =(u'^2 + v'^2)^{1/2}$. A Taylor length scale
can be defined as $\lambda=q'/\omega'$.

The flow is initialised with a fixed spectrum and random Fourier phases, and allowed to
evolve until structures have established themselves (figure \ref{fig:tur2d}). To investigate
the scaling of the different quantities, two initial reference times are used, $\omega'_0
t_{11}=11$ and $\omega'_0 t_{26}=26$, where the subindex denotes the moment at which a
variable is measured. As the simulation proceeds, vorticity decays to
$\omega'_{11}/\omega'_0=0.67$ and $\omega'_{26}/\omega'_0=0.47$. The kinetic energy
decreases more slowly, and the Taylor scale grows from $\lambda_0/L=0.03$ to
$\lambda_{11}/L=0.04$ and $\lambda_{26}/L=0.05$. Comparing these values with the vorticity
fields in figures \ref{fig:tur2d}(b,c) and with the spectra in figure \ref{fig:tur2d}(a),
$\lambda$ corresponds to the thickness of the thinnest vorticity filaments,
and is a few times smaller than the structures containing most of the enstrophy. The
kinematic viscosity $\nu$ is set to $q'_0L/\nu= 2500\,(q'_{11}\lambda_{11}/\nu
\approx 90)$, and the simulation uses $85^2$ complex Fourier modes. The numerical
resolution expressed in terms of Fourier modes improves from $\Delta x/\lambda =0.14$ at
$t_{11}$, to 0.11 at $t_{26}$, and there are about eight grid points across the smallest
vorticity structures during the time of the experiment. The energy spectrum included
in figure \ref{fig:tur2d}(a) shows that the scale disparity between energy and enstrophy is
approximately half a decade.
%

As shown in figure \ref{fig:tur2d}(b,c), the field is divided in a regular grid for the
modification step PL\ref{ploff:2}, and subsets of a given number of cells of this grid are
tested by zeroing their vorticity at time $T=0$. This is convenient in
two-dimensional flow because vorticity is not subject to continuity constraints, but spatial
periodicity requires that $\bra\omega\ket=0$. This is enforced by adding an appropriate
constant, $\omega_{00}$, to the whole flow field,
\beq
\widetilde{\omega}= \left\{\begin{array}{ll}%
\omega_{00} & \mbox{in}\, A, \\ 
\omega_{ref}+\omega_{00}& \mbox{otherwise}.
\end{array}\right.
\la{eq:delome1}
\eeq
To test the effect of cell size, we use grids with $N=16\, (4\times 4)$, $N=36\,
(6\times 6)$, and $N=100\, (10\times 10)$.
The effect of changing the modification strategy is discussed in \S\ref{sec:delete}.

\begin{figure}
\vspace*{2mm}%
\centerline{\includegraphics[width=.95\textwidth,clip]{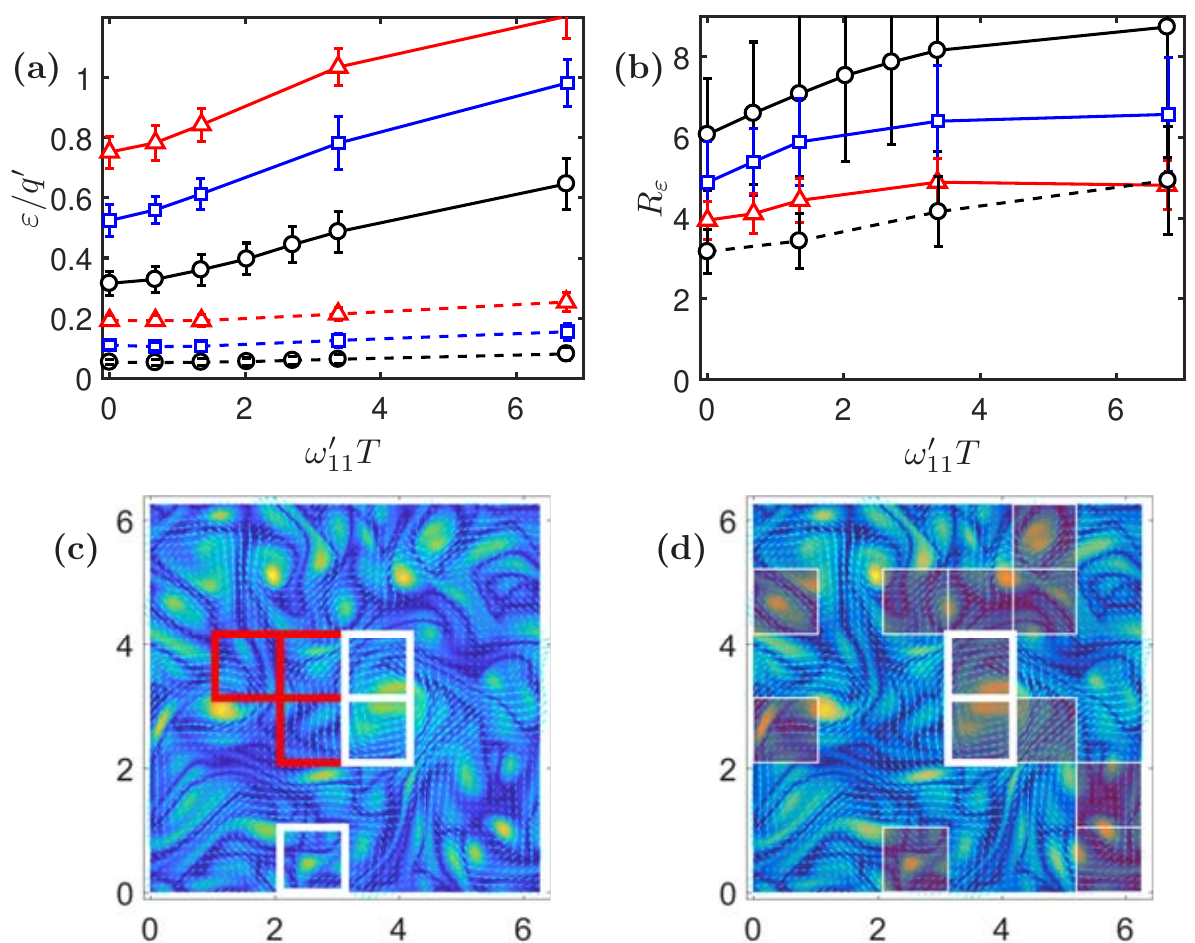}}%
%
\caption{%
(a) Evolution of the largest (solid) and smallest (dashed) deviation norm after deleting subsets of: \circle,
$n=1$; \squar, 3; \trian, 7 cells, out of 36. Bars are the standard deviation of the
variability over 48 initial conditions (192 for $n=1$).
(b) As in (a), for the significance ratio $R_\diss=\diss_{max}/\diss_{min}$. The dashed line
with circles is $n=1$, with cells deleted by zeroing the velocity instead of the vorticity, as in \r{eq:delu}. 
(c) Outlined in white is the most significant subset of $n=3$ cells out of $N=36$, defined at
$\omega'_{11} (t-t_{11})= 3.4$, and overlaid on the vorticity magnitude. The least significant
subset is outlined in red. (d) As in (c). Darkened cells belong to at least one of the ten
most significant subsets. The two cells heavily outlined in white belong to all of them.
}
\label{fig:ploff3}
\end{figure}

Figure \ref{fig:ploff3}(a) shows the evolution of the perturbation magnitude for the most
and least significant subsets of $n$ cells in a $N=36$ test grid. It suggests that
significance is largely a matter of representation, and that the deviation after some time
is roughly proportional to the representation difference at $T=0$, but figure
\ref{fig:ploff3}(b) tells a slightly different story. It shows that the ratio $R_\diss$
between the largest and the smallest deviation norm from subsets of a given size is
relatively large, and that it increases with the evolution time at which it is measured,
at least up to $\omega' T\approx 4$. The former proves that there are flow regions
that are substantially more significant than others, while the latter shows that
significance to the flow evolution involves more than initial representativeness. There are
flow regions that have a stronger effect on the evolution of the flow than what could be
expected from the energy that they contain. Figure \ref{fig:ploff3}(b) also shows that the
subsets which are most effective in determining significance are the smallest ones, $n=1$.
This is reasonable, at least on average, because there would be in general a most and a
least significant grid point in each flow field, and modifying more grid points dilutes
their significance.
  
Figure \ref{fig:ploff3}(c) shows an example of the most and the least significant subsets
with $n=3$ and $N=36$, measured at $\omega_{11} T=3.4$, and illustrates the tendency of even
these small subsets to be formed by collections of smaller ones. It turns out that there are
often several subsets with very similar significance. For the case in figures
\ref{fig:ploff3}(c,d), $\diss_{max}/\diss_{min}=5.45$, but the ratio between the lowest and
the tenth lowest deviation (out of approximately 7000 possible combinations) is about 1.1,
and that between the highest and the tenth highest deviation is 1.07. This is a common
feature of optimisation over many degrees of freedom \citep{LecunEtal2015} but, at least in
this particular case, it appears to be connected with the sharing of significant
sub-structures among subsets. Figure \ref{fig:ploff3}(d) shows the union of the ten most
significant subsets. Their component cells are scattered across the flow, but the two cells
heavily outlined in white are present in all the subsets, suggesting that the search
algorithm identifies individual significant cells and forms with them larger connected
composite substructures.

\begin{figure}
\vspace*{2mm}%
\centerline{\includegraphics[width=.85\textwidth,clip]{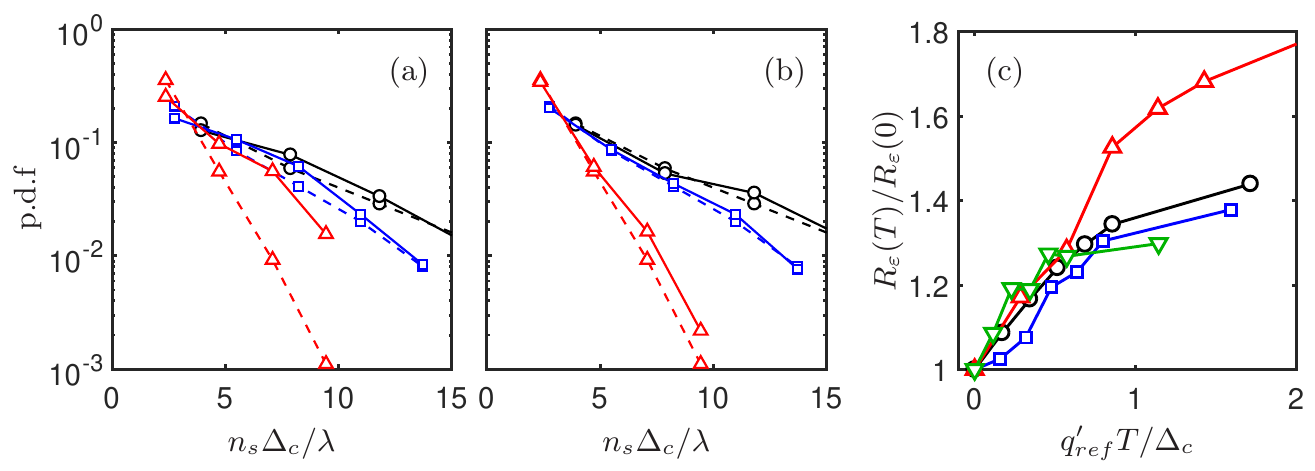}}%
%
%
\caption{%
(a) Size distribution of the connected components in the ten most significant subsets with $n=5$ (solid),
compared with the equivalent distribution for random sets of cells (dashed).
(b) As in (a), for the ten least significant subsets.
(c) Significance ratio measured at $T=t-t_{ref}$,  normalised with its value at $T=0$. $n=1$.
Time is scaled with the eddy turnover at the cell size, $\Delta_c$. 
In all figures,
\trian, $(N, t_{ref}, \Delta_c/\lambda)=(100, t_{11}, 2.4)$; 
\squar, $(36, t_{26},2.8)$; 
\circle, $(36, t_{11},4)$; 
\dtrian, $(16,t_{11},6)$.  
}
\label{fig:ploff1}
\end{figure}

Connected substructures of more than one cell form by chance even in Poisson-distributed
sets of cells, and how different are significant sets from random ones can be
characterised by comparing the probability distribution of the number, $n_s$, of cells in their connected
substructures. Defining connectivity by the eight nearest neighbours, this is done in figure
\ref{fig:ploff1}(a,b) for the most and least significant subsets, respectively. Their
behaviour is different. Figure \ref{fig:ploff1}(a) shows that the connected substructures of the
most significant sets tend to be larger than random ones. This difference scales best with
the ratio of the length $n_s \Delta_c$ to the Taylor microscale, where $\Delta_c=L/N^{1/2}$
is the cell size, and is maximum for intermediate sizes of the order of the diameter of
individual vortices. Not shown in the figure is another interesting difference. If the
inertial tensor of a connected substructure is computed by substituting cells by unit point
masses, an elongation can be defined by the ratio of its two principal moments of inertia.
Random sets tend to be approximately circular, with elongations close to unity, but the most
significant subsets are more elongated. This is especially true for the grid which is finest
with respect to the Taylor scale, which is also the one in which the distribution of
component size differs most from the random case (triangles in figure
\ref{fig:ploff1}a). The simplest interpretation is that finer grids are able to adapt the
shape and size of significant sets to the shape of the strained vortices seen in figure
\ref{fig:tur2d}. The tendency to group cells into linear filaments is also reflected in the
empirically best scaling length for the abscissae in figure \ref{fig:ploff1}(a,b), which is
the linear length of $n_s$ elements of size $\Delta_c$, rather than the diameter,
$n_s^{1/2}\Delta_c$, of a blob with the same number of elements.

Figure \ref{fig:ploff1}(b) show that most of these observations do not apply to the least
significant subsets, whose distribution of connected substructures is
essentially the same as for Poisson sets. It is also true that their elongations are much
closer to unity than for the most significant subsets. The comparison suggests that, while the
most significant sets reflect the structure of the flow, with specific geometries and
sizes, the least significant ones tend to be random collections of backgrounds cells.

The effect of deleting cells is not simply additive. Different cells tend to cancel each
other, and the effect of a small number of cells is weaker than the sum of its individual
components. Additivity improves for longer test times, presumably because the effect of
different components decorrelates. For example, in the cases in figure \ref{fig:ploff3}, the
effect at $T=0$ of deleting the most significant subset of three cells is approximately 1.6 times
weaker than the sum of the effects of the three most significant single cells, but that
factor is 1.3 for $\omega_{11}T=3.4$, and almost unity for $\omega_{11}T=6.7$.

Figure \ref{fig:ploff1}(c) shows the temporal evolution of the significance ratio with the
test time $T$ for several test cases and $n=1$. It follows from figure \ref{fig:ploff3}(b)
that the evolution of this ratio is relatively independent of $n$, although its absolute
value depends strongly on the size of the subset and on other details of the initial
conditions. To compensate for this dependence, the ratio is normalised in figure
\ref{fig:ploff1}(b) with its value at $T=0$. The time in this figure scales best with the turnover time
associated with the cell size, $\Delta_c$, rather than with the blob diameter,
$n_s^{1/2}\Delta_c$, specially for times of the order of a turnover. This suggests that the
labelling procedure isolates features of the order of $\Delta_c$, implying again thin
filaments rather than blobs.

\subsection{The effect of different modification strategies}\la{sec:delete}

Up to now we have discussed tests in which the whole flow field is modified in step
PL\ref{ploff:2} of the labelling algorithm, following \r{eq:delome1}. A different option
would be
\beq
\widetilde{\omega}= \left\{\begin{array}{ll}%
\omega_{00} & \mbox{in}\, A, \\ 
\omega_{ref}& \mbox{otherwise},
\end{array}\right.
\la{eq:delome2}
\eeq
where only the test subset is changed. This was found to make no appreciable difference
with respect to \r{eq:delome1}, most probably because $\omega_{00}$ is typically much
smaller than $\omega'$.

A more interesting possibility is to zero the velocity field within the chosen cells,
in which case continuity has to be taken into account because streamlines
are truncated at the edge of the cell. This option was also tested, projecting the modified
field onto the solenoidal subspace by means of a pressure-like gradient,
\beq
\vec{\widetilde{u}}= \left\{\begin{array}{ll}%
\nabla p & \mbox{in}\, A, \\ 
\vec{u}_{ref}+\nabla p & \mbox{otherwise},
\end{array}\right.
\la{eq:delu}
\eeq
where $p$ is adjusted so that 
$\nabla\cdot\vec{\widetilde{u}}=0$. It turns out that this procedure is less effective than
\r{eq:delome1}. The resulting significance ratio for $n=1$ is included as a dashed
line with circles in figure \ref{fig:ploff3}(b), and is about twice weaker than
the comparable solid line for \r{eq:delome1}. The reason is that zeroing the velocity within
a cell does not modify the surrounding cells, and, in particular, it does not change the
circulation around the cell being modified. The vorticity originally contained in the cell
is merely displaced to a sheet along its edge, which the irrotational 
gradient added in \r{eq:delu} cannot modify. Therefore, zeroing the velocity is equivalent
to the less effective process of rearranging the vorticity, instead of deleting it.
 
This option is not explored further in this paper, although the results are used in the next
section, but an equivalent choice has to be faced in generalising the present procedure to
three-dimensional flows. Not only are the length scales of velocity and vorticity typically
farther apart in three dimensions than in two, but vorticity itself is subject to a
solenoidality constraint.

\subsection{Properties of the significant regions}\la{sec:vortices}

\begin{figure} 
\vspace*{2mm}%
\centerline{\includegraphics[width=.85\textwidth,clip]{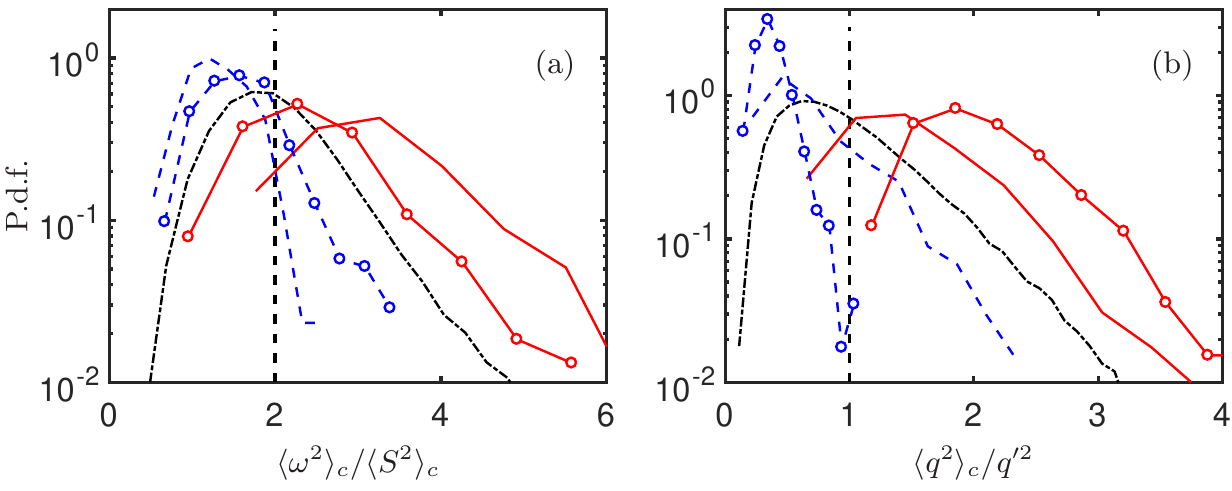}}%
%
%
\caption{%
(a) P.d.f.s of the ratio of mean vorticity to mean rate of strain over individual significant cells, for
192 initial conditions with $n=1,\, N=36,\, \omega'_{11}(t-t_{11}) =3.4$. The vertical
dashed line is the global average that separates vortex- from strain-dominated regions,
$\bra\omega^2\ket/\bra S^2\ket=2$. \solid, Ten most significant cells for each initial
condition; \dashed, ten least significant cells; \chndot, random cells. 
Lines without symbols are initialised using \r{eq:delome1}. Those with symbols use \r{eq:delu}.
(b) As in (a), for the kinetic energy, $q^2=u^2+v^2$.
}
\label{fig:oms}
\end{figure}

Although the main result of the previous analysis is the demonstration that there
are subsets of the flow with very different dynamical significance, it also allows us to
determine the properties of the most and least significant ones. Figure \ref{fig:oms}(a)
displays the p.d.f. of the ratio of the enstrophy to the magnitude of the strain rate,
averaged over the individual cells that have been labelled as most and least significant,
compared to the p.d.f. compiled over randomly chosen cells. When this ratio is taken over
the whole box it is kinematically true that $\bra\omega^2\ket/\bra S^2\ket=2$, which
separates vorticity-dominated regions from those dominated by the rate of strain.
The figure includes subsets modified by zeroing the vorticity, as in \r{eq:delome1},
together with others obtained by zeroing the velocity, as in \r{eq:delu}. In both cases, the most
significant sets are vortices, and the least significant ones are the strain-dominated gaps
between them, although the distinction is clearest for the cases that initially manipulate the
vorticity. This agrees with the qualitative impression from figure \ref{fig:ploff3}(c), and
with previous intuition \citep{mcwilliams90b}, and applies to all combinations of cell and
subset size. Figure \ref{fig:oms}(b) shows similar results for the kinetic energy. The
tendency is also for highly significant cells to be energetic, although, in this case, it is
clearest for the case that initially manipulate the velocity. The tendency for significance
to follow representativeness is to be expected, but it is reassuring that similar
conclusions are reached from cross-manipulation, in which the behaviour of a variable is
obtained from the modification of a different one.

\section{Discussion and conclusions}\la{sec:conc}

We have shown that fully automatic comparison of simultaneous undisturbed and modified
simulations can be used to segment a flow into regions properly labelled as more or less
significant from the point of view of flow dynamics. Moreover, significance can be
quantified as the growth of the perturbation after some time, and because regions are
labelled individually, their properties can be statistically identified. In our application
to two-dimensional turbulence, we find that the significant structures are vortices or
vortex filaments, while the interstitial strain-dominated regions are less significant. We
also show that significance is different from representativeness
at a fixed time. Some flow features are more dynamically significant than what could be
expected from their instantaneous energy content, and it is reassuring that tests
based on the modification of different variables give similar results. Moreover, the results
of the segmentation can be used to characterise the time scales and geometries of significant
features. It follows from figure \ref{fig:ploff1} that the key quantity controlling the
scales being observed is the size, $\Delta_c$, of the test cells.

The problem is posed as a computer `game' in which the machine tests the effect of
different modifications of the initial conditions, and selects those that maximise or minimise it. The
main difficulty of the procedure is cost, due to the large number of initial conditions and
modifications that have to be tested, but it can benefit from the body of recent
research on similar problems in data analytics and artificial intelligence. Even simple
approximations help. While the purpose of the present paper is to test the
strategy rather than to optimise processing time, considerable savings are achieved by 
substituting exhaustive searches by reasonable approximations, and the results presented above
only represent a few thousand hours of computer time. In addition, the tests of different initial
conditions are essentially independent, and can be trivially parallelised.

In retrospect, it should be clear that the problem of structure identification has only been
partially automatised. In true machine learning, the result of the game is fed back into the
initial conditions to optimise the result. In our experiment, this `back propagation' step
is left to the researcher, who evaluates the result and proposes modifications. 
The choice of test grids, initial flow conditions, observation times and deletion
procedures discussed in \S\ref{sec:tur2d} was motivated in this way. The automatic part of
the procedure described here can thus be seen as a fast and relatively exhaustive `forward' search
over experiments, resulting in the automatic labelling of flow subsets in terms of
their significance.

Its main utility is probably its relative independence from preconceived ideas, which may
mislead the researcher into confusing dynamical significance with representation of a
particular quantity. While it would have been surprising if two-dimensional turbulence had
been found to be non-intermittent, or if its significant eddies had been found to be other
than vortices, the situation in more complex flows is, at least to the author, less clear.
For example, while it is probable that we have identified by now the most significant types
of structures of wall turbulence, especially near the wall, it is not impossible that we
have missed some, and the situation is even less certain in homogeneous three-dimensional
turbulence. Generalising the present results to these cases would require additional
work, especially in the choice of initial perturbations, but, at least for moderate
Reynolds numbers, both problems should be accessible to the current generation of computers.

\vspace{2ex}
This work was supported by the European Research Council under the Coturb grant
ERC-2014.AdG-669505. 
%
\appendix\section{Search algorithms and operation estimates}\la{app:search}

An exhaustive search of all possible subsets is only practical in simple cases, but a
`pruned forest' approximation turned out to give nearly perfect results, although 1000 times faster.
The number of tests needed for a full search of $n$ elements from a set of $N$ cells is the
binomial coefficient ${N\choose n}$, which grows factorially with $N$ and $n$. In the
example used in figure \ref{fig:ploff3}, the number of tests for $N=6^2$ and $n=3$ is about
7000, but it grows to $3.7\times10^5$ for $n=5$, and to $8\times 10^6$ for $n=7$. The number
of tests required for a full search of the case $n=12$ has ten digits.

\begin{figure} 
\centerline{\includegraphics[width=.95\textwidth]{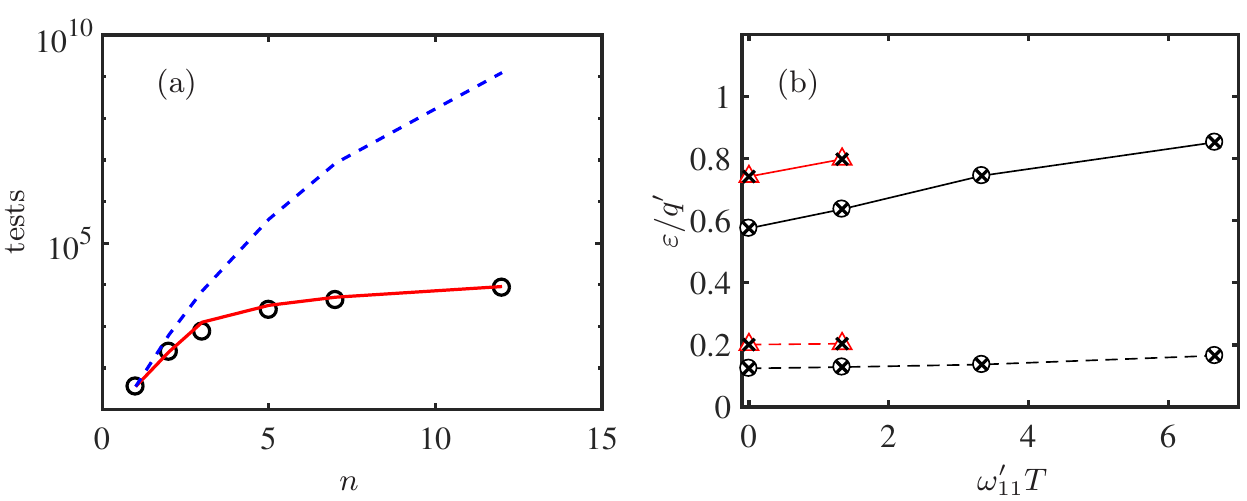}}%
%
%
\caption{(a) Number of tests required for: \dashed, full search; \circle, tree search;
\solid, estimate in \r{eq:tests}.
(b) Comparison of the results of the tree search (symbols), and the full search (crosses).
\circle, $n=3$; \trian, $n=7$. Solid lines are the highest deviation limit, and dashed ones are
the lowest. In all cases, $N=36$, the pruning size is $M=15$, and the branching number is
$b=3$. }
\la{fig:tests}
\end{figure}

Our basic algorithm for the construction of optimal subsets is a greedy decision
tree, in which a `best' cell is modified and tested first, and the subset is constructed by
incrementally adding the best of the remaining pool of cells. Such greedy algorithms are not
generally exact \citep{CorLeiRiv:90}. In the particular case of decision trees, the main
problem is to prevent them from getting stuck in local extrema, and a popular remedy is to
introduce noise by considering randomly initialised `forests' \citep{Breiman:01}, instead of
deterministic trees. In our case, where we are not so much interested in global optima as in
suboptimal classes, the randomness is introduced by keeping several possibilities after each
decision, successively pruning them to avoid exponential branching.

The  algorithm proceeds as follows:
\begin{enumerate}
\item Test the $N$ single cells, and populate the first level of the search forest with those
resulting in the $b$ smallest deviations.
\item\la{repeat} To construct the $j$-th level, extend each element in level $j-1$ with
one of the $N-j$ cells that have not been used, and retain the $b$ smallest deviations. After
all the $(j-1)$-th trunks have been extended in this way, prune the $j$-th forest level to a
maximum of the best $M$ non-repeating elements.
\item Repeat step \ref{repeat} until the final $n$-th level is reached.  
\end{enumerate}
The same procedure is performed for the largest deviations. It is clear from this description
that the maximum number of required tests is $O(MNn)$ as long as the pruning size is larger
than the branching ratio, $M>b$. In the text, $M=15$, $b=3$, and the number of tests needed
for the largest case mentioned above $(N=36,\, n=12)$ is approximately 9000, instead of the
$10^9$. A more precise estimate for the maximum number of tests is
\beq
N_T= 2M(n-2)(N-(n+1)/2) +N+ 2b(N-1),
\la{eq:tests}
\eeq
where the last term only applies to $n>1$, and the first one to $n>2$.  It is compared in figure \ref{fig:tests}(a) to actual experimental values.     

The results differ little in practice from the full search, at a much lower cost. Figure
\ref{fig:tests}(b) shows some results for a particular initial condition. Symbols are the
result of the tree search, and crosses those of the full search.
 

\end{document}